\documentclass[aps,pre,twocolumn,showpacs,preprintnumbers,amsmath,amssymb,superscriptaddress]{revtex4}
\usepackage{graphicx}
\usepackage{dcolumn}
\usepackage{color}
\usepackage{bm}

\begin{document}

\title{The competition between surface adsorption and folding of fibril-forming polypeptides}

\author{Ran Ni}
 \email{rannimail@gmail.com}
 \affiliation{%
 Van $'$t Hoff Institute for Molecular Sciences, Universiteit van Amsterdam, Science Park 904, 1098 XH Amsterdam, The Netherlands
 }%
 \affiliation{%
 Laboratory of Physical Chemistry and Colloid Science, Wageningen University, Dreijenplein 6, 6703 HB Wageningen, The Netherlands
 }%

\author{J. Mieke Kleijn}
\affiliation{%
 Laboratory of Physical Chemistry and Colloid Science, Wageningen University, Dreijenplein 6, 6703 HB Wageningen, The Netherlands
 }%

 \author{Sanne Abeln}
 \affiliation{%
Centre for Integrative Bioinformatics (IBIVU), Vrije Universiteit, De Boelelaan  1081A, 1081 HV Amsterdam, The Netherlands
 }%

\author{Martien A. Cohen Stuart}
\affiliation{%
 Laboratory of Physical Chemistry and Colloid Science, Wageningen University, Dreijenplein 6, 6703 HB Wageningen, The Netherlands
 }%

\author{Peter G. Bolhuis}
\affiliation{%
Van $'$t Hoff Institute for Molecular Sciences, Universiteit van Amsterdam, Science Park 904, 1098 XH Amsterdam, The Netherlands
}%

\begin{abstract}
Self-assembly of polypeptides into fibrillar structures can be initiated by planar surfaces that interact favorably with  
certain residues.  Using a coarse grained model, we systematically studied the folding and adsorption behavior of a $\beta$-roll forming polypeptide. We  
find that there are two different folding pathways depending on the temperature: (i) at low temperature, the polypeptide folds in solution into a $\beta$-roll before adsorbing onto the attractive surface, (ii) at higher temperature, the polypeptide first adsorbs in a disordered state, and folds while on the surface.  The folding temperature increases with increasing attraction, as the folded $\beta$-roll is stabilized by the surface. Surprisingly,  further increasing the attraction lowers the folding temperature again, as  strong attraction also stabilizes the adsorbed disordered state, which competes with folding of the polypeptide. Our results suggest that to enhance the folding, one should use a weakly attractive surface.  They also  explain the recent experimental observation of the non-monotonic effect of charge on the fibril formation on an oppositely charged surface [C. Charbonneau, {\em et al.}, ACS Nano, 8, 2328 (2014)].
\end{abstract}



\maketitle

\section{Introduction}

{
The interest in spontaneous fibril formation  by structural proteins derives not only from the link with neuro-degenerative diseases such as Alzheimer’s and Parkinson’s, which  have been traced to undesired amyloid fibril formation ~\cite{disease}, but also from the fact that  many natural and synthetic proteins form fibrils and hold promise for application as novel biomaterials~\cite{Zhang2003,bioapp1,bioapp2,fibexp,lennart2012}.  In particular, the stimuli-responsive properties of fibrils  have generated a strong interest in biomedical application\cite{Kopecek2012}.  Fibril formation  is usually kinetically controlled  and occurs via a nucleation-growth mechanism.  While this mechanism is often believed to involve  homogeneous nucleation,  fibril formation can also be surface-induced. Such a mechanism  has practical applications in industrial biosensors, in biotechnology, and in nanotechnology\cite{bioapp1,Zhang2013}.
Surface-mediated fibril formation occurs in  oligopeptides (typically around 20 residues long) \cite{Ku2008,Whitehouse2005,Yang2007,Yang2007a} 
and amyloid beta peptide \cite{Blackley2000,Zhu2002,McMasters2005,Moores2011}, but also for  silk-elastine-like polymers \cite{Hwang2009,celine2014}.
The surface-induced fibril self-assembly process can occur via several routes.  One involves a nucleation-and-growth process in solution, after which the preformed seed adsorbs on the surface, and continues to grow \cite{celine2014,Ku2008,Whitehouse2005,Yang2007,Yang2007a,Blackley2000,Zhu2002,Moores2011}. Another is by  direct adsorption of the single molecules on the surface\cite{Ku2008,McMasters2005,Hwang2009,celine2014}.
The morphology of surface-induced fibers depends on the protein concentration, the physicochemical surface properties, and environmental conditions such as temperature, pH, and ionic strength\cite{Zhang2013,Ku2008,Yang2007,Yang2007a,Moores2011,Hwang2009,celine2014}.
}

We focus here on an example of  designed biosynthetic peptide polymers based on silk-like and collagen-like sequences\cite{fibexp}. 
These polypeptides consist of three connected blocks, with the central one, inspired by sequences occurring in natural silk, a repeated octapeptide GAGAGAGX, where G and A are glycine and alanine, respectively, and X is a polar residue such as glutamate or histidine. This middle block carries at either end a proline-rich and rather hydrophilic sequence inspired by natural collagen, which does not form any secondary structures but stays a random coil in aqueous solution.
The silk-like block can fold into a $\beta$-roll structure as soon as the charge on the polar residue is removed \cite{schor2009,schor2011}. Such $\beta$-rolls have hydrophobic alanine rich faces, by which they self-assemble
into long and filamentous stacks \cite{schor2009,schor2011}. At sufficiently high concentrations, these filaments form dilute hydrogels which are promising candidates as a matrix for artificial tissue\cite{fibexp,lennart2012}.
The question that concerns us here is how this process starts. 
Kinetic experiments indicated that individual molecules in solution do not readily fold under the conditions of the experiments\cite{lennart2012,celine2014}, and that fibril formation
is governed by a nucleation-and-growth mechanism. In some cases (e.g., for X = histidine) homogeneous nucleation seems to occur, but in others (e.g., when X = glutamate) this does not happen~\cite{fibexp}. Recent experiments have shown that, in line with this, the presence of a surface which is weakly attractive to certain residues in the peptide sequence can promote the formation of fibril structures, but the effect is rather subtle and the underlying physics remains unclear~\cite{celine2014}. 
We have shown earlier that the formation of these fibrils is mainly triggered by the presence of a folded polypeptide, which then serves as a seed for the further growth of the long fibril~\citep{Ni2013prl}. 
The purpose of this work is to elucidate, by using computer simulations, whether or not a flat surface can indeed take the role of the seed, and how the interactions between surface and polypeptide residues influence this. Since the experiments \cite{celine2014} suggested that the polar residue X is attracted towards the surface, we pay specific attention to this residue.  To have an explicit model, we here choose X to be glutamate (E), but our conclusions should mutatis mutandis be valid for other choices.  

\section{Model}
Although in principle atomistic models can provide insights into the folding of polypeptides, such all-atom simulations are prohibitively expensive for the polymers considered here. 
Therefore coarse grained modelling is the method of choice for making progress in understanding the physics of protein folding~\cite{aggfibril,auerphasediag,review,schor2011,ckhall0,ckhall1,ckhall2}.
Here, we employ a coarse-grained polypeptide model in which each residue occupies a single site on a 3D cubic lattice, with all other sites considered as solvent~\cite{sanne2011,Ni2013prl}. In contrast to implicit solvent protein models, this model has been shown able to prevent artificial aggregation of proteins in their native state.
Each residue has a unit vector indicating the direction of its side chain.
Two essential features for describing folding, i.e., the formation of hydrogen bonds and the directionality of side chains, are correctly captured in this highly efficient model. The total potential energy of the system is given by
$       E = E_{aa} + E_{solvent} + E_{hb} + E_{steric},$
where $E_{aa}$ and $E_{solvent}$ are are interaction potentials between residues, and between a residue and solvent, respectively; the values of these have been obtained by comparison with experimental data as shown in Table ~\ref{intermatrix} (all interaction potentials in this work are in reduced units).
$E_{hb}$ is the potential energy of formed hydrogen bonds, and $E_{steric}$ represents the steric hindrance between consecutive residues in a polypeptide chain~\cite{sanne2013}.
Two amino acids in contact interact only when their side chains are either in parallel or pointing towards each other.
Similarly, interaction between a residue and solvent only exists when the side chain points to a solvent site. When a residue is not part of a turn in the backbone, it can adopt a  strand conformation depending on the side chains. 
Two residues in contact and both in the strand state can form a hydrogen bond with an energy $\epsilon_{hb} = -0.5$, when their side chains are aligned.
An energy penalty of $\epsilon_{s} = 0.55$ is employed to prevent the side chains of consecutive residues pointing in the same direction, thereby mimicking steric hindrance and restrictions in bond rotation.
Due to its small size, alanine in a $\beta$-strand environment behaves as a considerably more hydrophobic residue than it actually is; this effect is not well captured by the original parametrization of the potential. To compensate for this shortcoming, we vary the alanine-water interaction parameter to investigate its influence.
Configuration space is explored using a lattice Monte Carlo scheme with a classical set of moves~\cite{sanne2013}.

As lattice models in general are not expected to fold natural sequences, we designed a sequence to fold into the desired $\beta$-roll structure. As explained in Ref\cite{Ni2013prl} proper  folding of a  $\beta$-roll on a  cubic lattice demands a palindromic sequence and a 
anti-parallel sheet topology. 
We restrict the silk part to 80 residues to make the calculation tractable, for which 
the design procedure yielded the optimal sequence $(E(AI)_3RE(IA)_3R)_6$.  The replacement of glycine with isoleucine 
is not unrealistic, as glycines in $\beta$-sheets are more hydrophobic than the average glycine~\cite{Ni2013prl}.  
The extra (arginine) residue has been introduced to fit the $\beta$-roll structure on the lattice. 

\begin{table}[t]
\caption{Interaction matrix (in reduced units) for the residues in the used sequence~\cite{sanne2011}. Amino acids are denoted by their one-letter code (I = isoleucine, A = alanine, R = arginine, E = glutamate), and $w$ and wall denote the solvent (water) and the residues on the surface. The hydrophobicity of alanine is varied via the A-w interaction ($\epsilon_{A,w}$) as indicated.}\label{intermatrix}
\begin{tabular}{rrrrrrr}
        & I             & A             & R             & E             & w     & wall\\
\hline
 I      & -0.79 & -0.40 & 0.5   & 0.69  & 0.7 & 0\\
 A      &               & -0.34 & 0.49  & 0.77  & [0.01, 0.6] & 0\\
 R      &               &               & 0.43  & -0.6  & -0.57 & 0\\
 E      &               &               &               & 1.02  & -0.78 & [0.0, -3.0]\\
 w      &               &               &               &               & 0.0 &0\\
 wall      &               &               &               &               &   &0\\

 \hline
\end{tabular}
\end{table}

\begin{figure}[t]
        \includegraphics[width=0.45\textwidth]{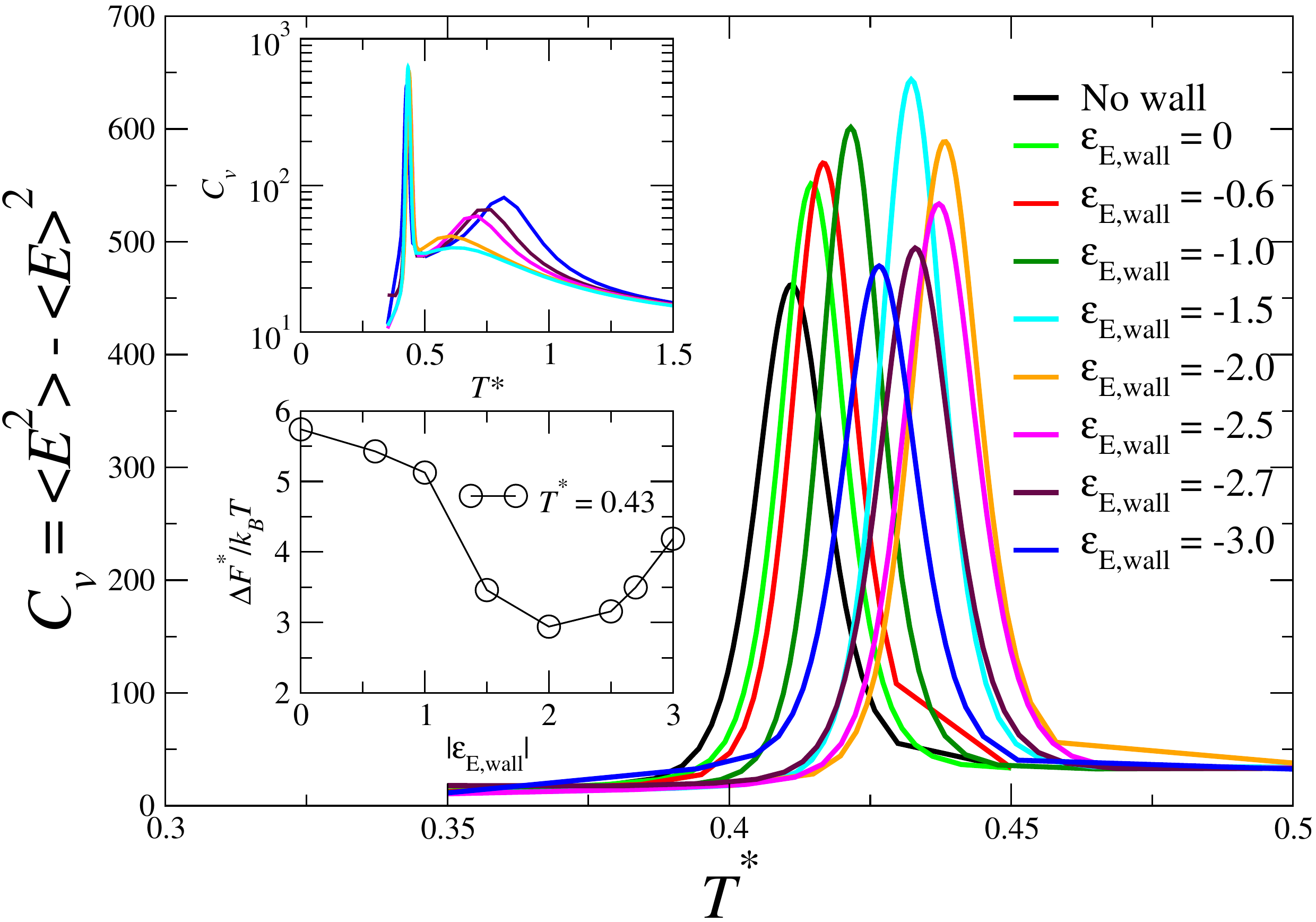}
        \caption{(color online) Heat capacity $C_v = \langle E^2 \rangle - \langle E \rangle^2$ as a function of reduced temperature $T^*$ for various strength of attraction between the wall and glutamate $\epsilon_{E,wall}$ around the folding temperature for a single polypeptide with alanine hydrophobicity $\epsilon_{A,w} = 0.6$. Inset: (Top) Heat capacity $C_v$ as a function of $T^*$ for the whole temperature range  including both folding and adsorption of the polypeptide. (Bottom) The free energy barrier height of peptide folding, $\Delta F^*/k_BT$, as a function of surface attraction strength $|\epsilon_{E,wall}|$ at a constant temperature $T^* = 0.43$.}\label{fig1}
\end{figure}

\begin{figure*}[t]
        \includegraphics[width=0.9\textwidth]{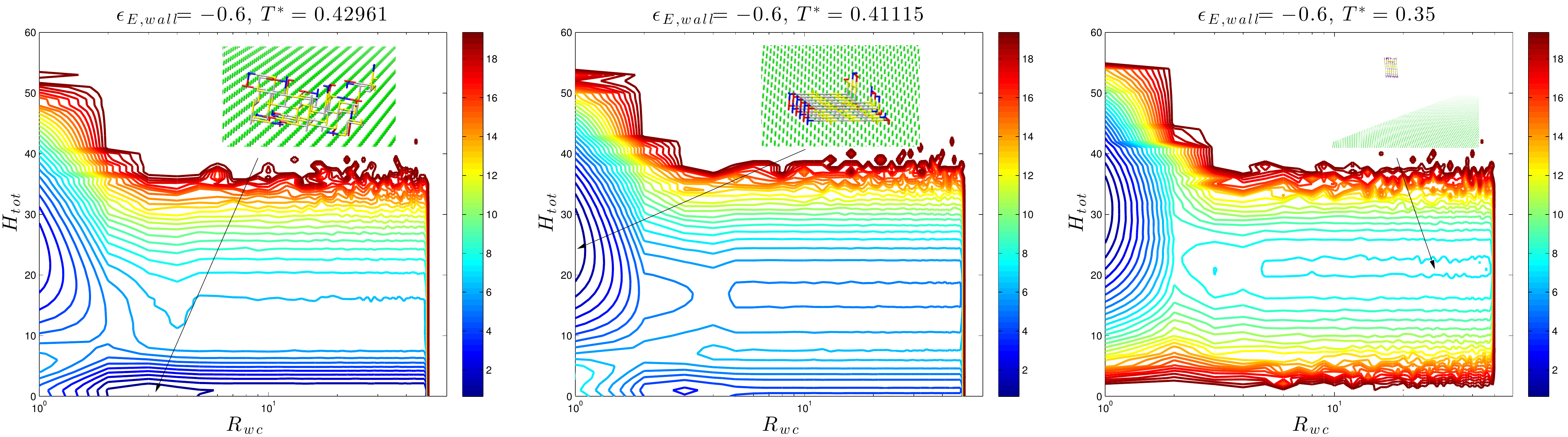}
        \includegraphics[width=0.9\textwidth]{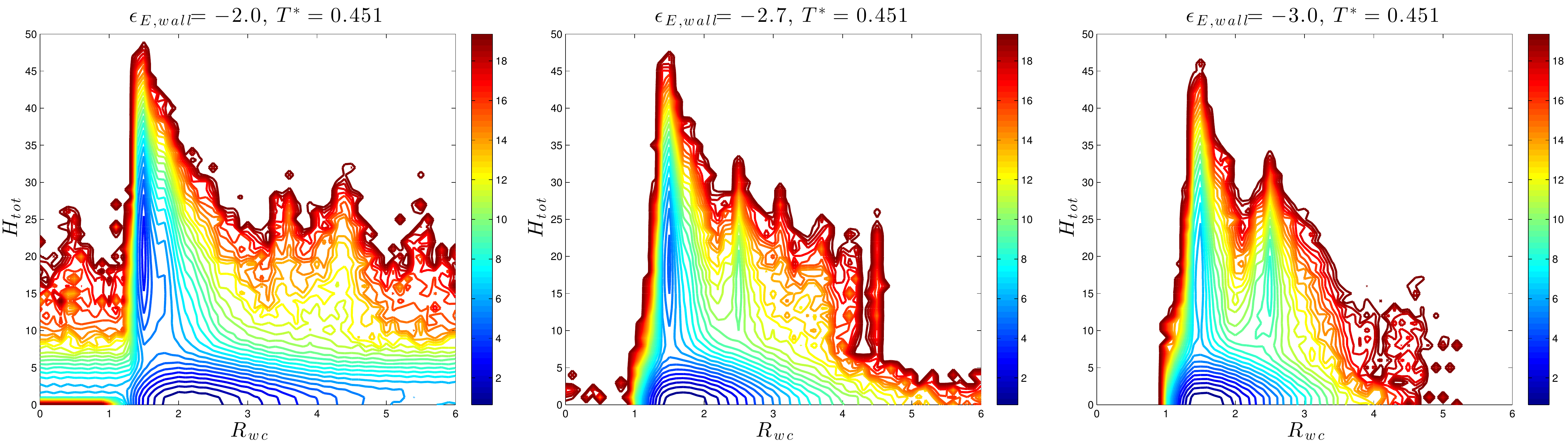}
        \caption{(color online) Top: Free energy landscapes as a function of the distance between the polypeptide  center of mass  and the wall, $R_{wc}$, and the total number of hydrogen bonds, $H_{tot}$, for  a single polypeptide with the alanine hydrophobicity $\epsilon_{A,w} = 0.6$, and wall attraction $\epsilon_{E,wall} = -0.6$ at different temperatures. Insets:  Snapshots of a typical configuration corresponding to the local minima on the free energy landscapes indicated by the arrows (from left to right): an unfolded molecule, a folded molecule at the surface, and a folded one in solution. Residue color coding: alanine(A)=yellow, isoleucine(I)=white, glutamate(E)=red, arginine(R)=blue, and wall=green. Bottom: Free energy landscape as a function of $R_{wc}$ and $H_{tot}$ for the system containing a single polypeptide with $\epsilon_{A,w} = 0.6$, and various wall attraction $\epsilon_{E,wall}$ at the temperature $T^* = 0.451$.}\label{fig2}
\end{figure*}

We performed replica exchange Monte Carlo (REMC) simulations for one polypeptide chain with sequence  $(E(AI)_3RE(IA)_3R)_6$ in a simulation box of $100\times 100 \times 100$ lattice sites with periodic boundary conditions in $x$ and $y$ directions. 
 To mimic the experimental situation, in which the X (here, E) residue is electrostatically attracted to the surface 
we put two parallel walls on opposite sides of the box in the z direction consisting of sites which exclusively attract E with a variable strength $\epsilon_{E,wall} $ 
(see Table \ref{intermatrix}). We ensure that the size of the simulation box is large enough not to influence the folding of the polypeptide. 
Each REMC simulation consisted of 48 replicas with a (reduced) temperature distribution around the transition temperature, which is optimized by a feedback-optimization algorithm ~\cite{fbo}. 
For each replica, we performed $6 \times 10^{10}$ MC cycles with an exchange attempt every 1000 cycles. The first $10^{10}$ moves were for equilibration.
 Employing the virtual-move parallel tempering method during the production~\cite{vmpt} ensured optimal use of the simulation data.

\section{Results}
We first simulate the folding of the polypeptide with a relatively strong alanine hydrophobicity $\epsilon_{A,w} = 0.6$, which is close to the experimental situation~\citep{Ni2013prl}. The heat capacity $C_v$ calculated from the energy fluctuations, as a function of the reduced temperature $T^*$  is shown in Fig.~\ref{fig1}  for different surface attraction strengths. 
Even when the surface is not attractive at all, i.e. at $\epsilon_{E,wall} = 0$, the folding temperature is somewhat higher than the bulk value, implying that the surface stabilizes the folded structure.
This stability follows from the hydrophobicity of the outside surface of the folded $\beta$-roll structure (exposing mostly alanine),  which 
prefers 
the surface. With increasing attraction strength, from $|\epsilon_{E,wall}| = 0$ to 2.0, the folding temperature rises. 
The corresponding folding free energy barrier height is shown in the inset of Fig.~\ref{fig1} for  
a temperature  $T^* = 0.43$ at which the polypeptide cannot spontaneously fold 
in the bulk solution. The barrier   dramatically decreases with attraction to the wall, indicating  the 
 attraction  promotes the folding of the polypeptide.

Fig.~\ref{fig2} top shows typical free energy landscapes
for $\epsilon_{E,wall}=- 0.6$  at different temperatures, as a function of the distance between the mass center of the polypeptide and the surface,  $R_{wc}$, and the total number of hydrogen bonds formed in the system, $H_{tot}$.
At low temperature $T^* = 0.35$ 
the stable phase of the system is the folded $\beta$-roll structure,  represented by a global minimum at $(R_{wc} \approx 1, H_{tot}. \approx 30)$.
Another  local minimum at $(R_{wc} \approx 30, H_{tot} \approx 20)$ corresponds to a folded $\beta$-roll floating at a distance from the surface. 
At this temperature the polypeptide is likely to first fold in solution before adsorbing onto the attractive surface. 
At $T^* = 0.42961$ (Fig.~\ref{fig2}, top left), just above the folding transition,   the global minimum is at
$(R_{wc} \approx 3, H_{tot} \approx 0)$, corresponding to a {\em disordered} polypeptide adsorbed on the surface.
A channel appears that connects the disordered and the folded structures on the surface.
At $T^* = 0.41115$, slightly below the wall-induced folding temperature $T_f = 0.4166$, but still above the bulk folding temperature $T_f^{bulk} = 0.406$ (Fig.~\ref{fig2},top middle panel),
the adsorbed $\beta$-roll structure becomes the stable state of the system.
This strongly suggests that the adsorbed disordered polypeptide folds while in contact with the surface. The presence of the attractive surface thus not only stabilizes the $\beta$-roll, but, as a first step towards folding, also brings the disordered structure towards the surface. 
For this particular $\beta$-roll structure, 
only half of the 
octapeptide strands, and hence half of the E residues, are in  direct contact with the surface.
This is similar to the structure found in atomistic simulations where the E residues 
were pointing in opposite directions on the two sides of the 
folded $\beta$-roll~\cite{schor2011}.


Upon increasing the surface attraction strength
$|\epsilon_{E,wall}| = 2.0$ to 3.0,  intriguingly, 
the folding temperature
moves down, and the free energy barrier for folding increases again.
Simultaneously, a new high temperature peak appears in the heat capacity. This peak corresponds to the adsorption/desorption transition of the polypeptide, and shifts to higher temperatures with increasing surface attraction strength due to the enhanced stability of the adsorbed phase. 
To see why the folding temperature moves down, we again plot free energy landscapes as a  function of $R_{wc}$ and $H_{tot}$  in Fig.~\ref{fig2} (bottom) for  a fixed temperature ($T^* = 0.451$) above the folding transition for varying surface attractions from $|\epsilon_{E,wall}| = 2.0$ to 3.0.
In these diagrams the two local minima, located at $(R_{wc} \approx 1.5, H_{tot} \approx 20)$ and $(R_{wc} \approx 2, H_{tot} \approx 0)$, correspond to the adsorbed folded and disordered state, respectively. 
The dissolved, non-adsorbed state is now completely suppressed. 
While at $\epsilon_{E,wall} = -2.0$ the two states differ only slightly in free energy, 
increasing the attraction stabilizes the disordered state, and enlarges this difference to as much as $10k_BT$ for $\epsilon_{E,wall} = -3.0$.  
Along with this change, the average peptide-wall distance $R_{wc}$ 
decreases as the chain flattens out on the surface.
As in the folded structure only one half of the glutamate residues are in direct contact with the surface, the  remaining glutamate residues also tend to stick to the surface when the attraction becomes too strong, but this can only happen if the $\beta$-roll structure is destroyed. Such behavior is also observed in simulations of small proteins next to an attractive wall \cite{wolynes1994,shea2004,shea2006,shea2010,Marino2012}.
Hence, the effect of attraction on the efficiency of surface-induced folding is non-monotonic, because of the specific topology of the folded structure. We note that as the surface adsorbed  $\beta$-roll 
exposes one alanine-rich face toward the solution, it likely will act as a nucleus for fibril  growth by favorable interaction between two such faces ~\citep{Ni2013prl}.

\begin{figure}[t]
        \includegraphics[width=0.45\textwidth]{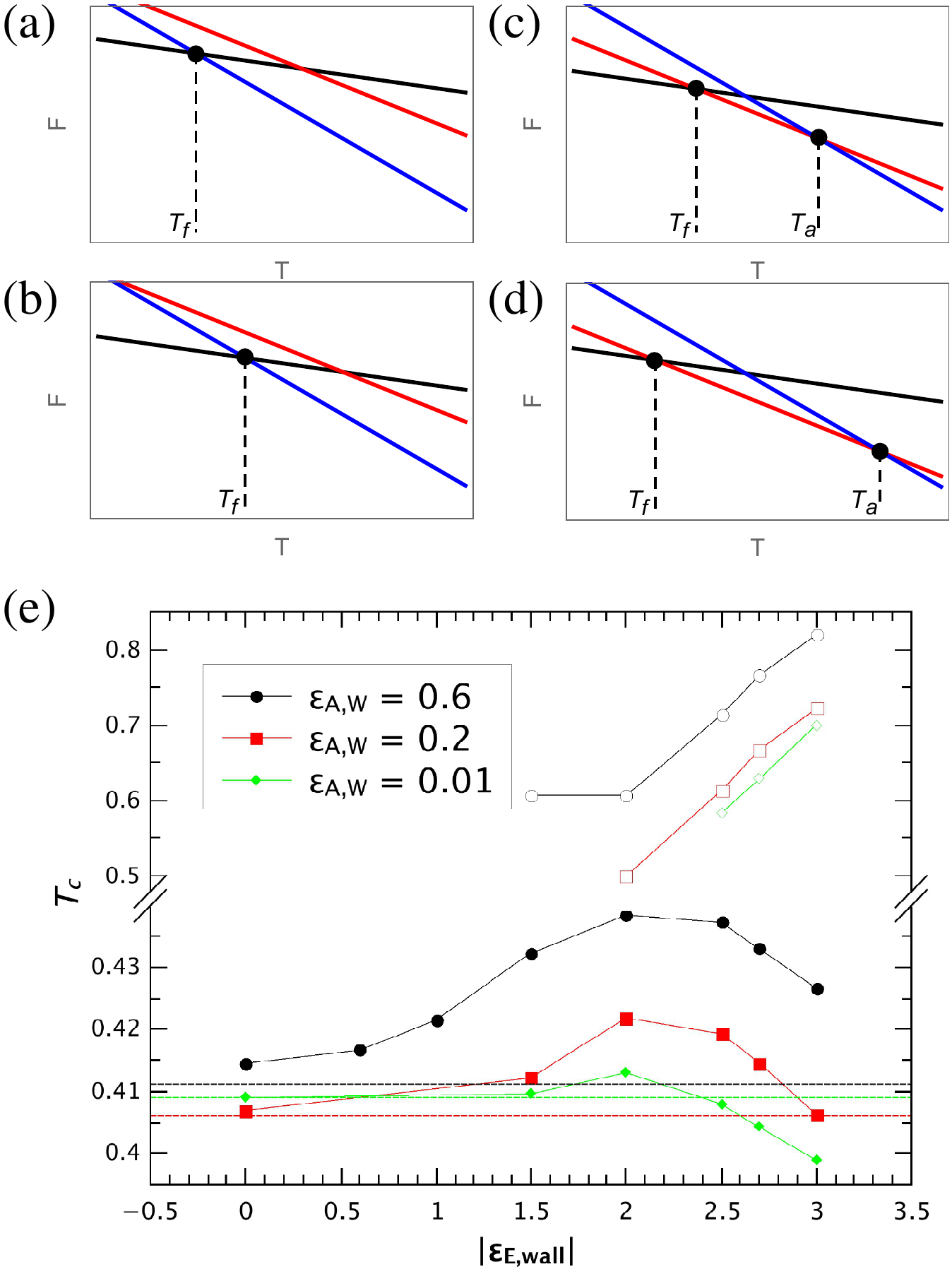}
        \caption{(color online) (a-d) The schematic illustration of the free energies of three  states of the system $F$ as a function of temperature $T$ with increasing wall attraction from (a) to (d). Black, red, and blue lines are the free energies for folded state, disorder state on the wall and the disorder state in the bulk solution, respectively, and $T_f$ and $T_a$ are the folding and adsorption temperatures, respectively. (e) Critical temperature, $T_c$, for the transitions, i.e. folding (filled symbols) and adsorption (open symbols), as a function of the wall attraction $|\epsilon_{E,wall}|$ for the polypeptide with various alanine hydrophobicities $\epsilon_{A,w}$. The horizontal dashed lines indicate the folding temperature of polypeptide in bulk solution.}\label{fig4}
\end{figure}

The folding mechanism of the polypeptide on an attractive surface is explained with schematic plots of the free energy vs.~temperature
 in Fig.~\ref{fig4}a-d. For low surface attraction, the polypeptide folds in the bulk solution when decreasing the temperature of the system, which occurs as a direct transition from the disordered state to folded state at $T_f$ (Fig.~\ref{fig4}a). With increasing surface attraction, the folded state is stabilized, the free energy of the folded state decreases, and the folding temperature decreases (Fig.~\ref{fig4}b). A very strong surface attraction also stabilizes a disordered polypeptide on the surface, which splits the folding of polypeptide into  two steps: with decreasing temperature, the disordered polypeptide first adsorbs onto the surface at temperature $T_a$, before folding on the surface at $T_f$ (Fig.~\ref{fig4}c). 
Stronger surface attraction increases the adsorption temperature $T_a$ and lowers the folding temperature $T_f$ (Fig.~\ref{fig4}d).
Fig.~\ref{fig4}e  plots the two transition temperatures $T_a$  and $T_f$, 
as a function of surface attraction at various alanine hydrophobicities $\epsilon_{A,w}$. 
For all three cases, the surface effect is non-monotonic, with $T_f$ showing a maximum around $ \epsilon_{E,wall} = -2$ (lower curves) while the adsorption temperatures increase monotonically (upper curves). 
The mechanism seems to be independent of the hydrophobicity of the outer face, which is crucial for the stacking of the $\beta$-roll
~\citep{Ni2013prl}. 
{We note that  this explanation implies that there is a certain value for $\epsilon_{A,w}$  at which the disordered polypeptide just barely becomes stable. This is similar to the situation in which a liquid phase becomes stable, as a function of the pressure in a simple liquid phase diagram.  While in principle this special  value for $\epsilon_{A,w}$  does exist, it is hard to measure, even in our lattice model, due to the large error bars connected to the estimating the transition temperature.}

\section{Conclusions}
In conclusion, by performing computer simulations of a coarse-grained lattice model polypeptide, we obtain mechanistic insights into the effect of surface attraction on the folding of a single disordered polypeptide with a silk-like sequence. We find that 
increasing attraction between surface and polar residue stabilizes folding, i.e. the folding temperature increases, and the free energy barrier for the folding of polypeptide decreases. A pathway for surface-induced folding opens up along which the surface first captures the disordered chain which then can pass over to the folded state, even above the bulk folding temperature. In contrast, at lower temperature, folding can first occur in solution after which the folded structure adsorbs to the wall. 
At strong
surface attraction 
the polypeptide tends to flatten out entirely, rendering the folded structure unstable with respect to the disordered state, with a higher folding barrier.

Our results suggest, therefore, that in order to promote surface-induced folding to seed the hierarchical self-assembly of protein fibrils, one should experimentally operate in a window of relatively weak binding where the chain flattening does not occur. 
{
We stress that these findings are generic, and should translate to other protein systems.
}

\begin{acknowledgments}
We acknowledge financial support from the European Research Council through Advanced Grant 267254 (BioMate). 
This work is part of the research programme VICI 700.58.442, which is financed by the Netherlands Organization for Scientific Research (NWO).
This work has been supported by NWO VENI grants with the contract number 680-47-441 (R.N.) and 722-011-009 (S.A.).
\end{acknowledgments}

\bibliography{ref}

\end{document}